# Adjoint Approach to Beam Optics Sensitivity Based on Hamiltonian Particle Dynamics

Thomas M. Antonsen Jr.[i], David Chernin and John Petillo

Leidos, Inc., Billerica, MA 01821

**Abstract**

We develop a sensitivity function for the design of electron optics using an adjoint approach based on a form of reciprocity implicit in Hamilton's equations of motion. The sensitivity function, which is computed with a small number of time-reversed runs of a beam optics code, allows for the determination of the effects on specific beam quality figures of merit of small, but arbitrary changes in electrode potentials, positions and shapes, and in magnet strengths and locations. The sensitivity function can thus be used in an optimization cycle of a focusing system's design, and/or to predict the sensitivity of a particular design to manufacturing, assembly, and alignment errors.

1. **Introduction**

   The design and optimization of complex devices and instrumentation is common to all fields of science and engineering. The design of a particular device generally depends on many parameters, and an optimum design must be found in a large, potentially infinite, dimensional parameter space. A powerful tool useful in the design process is a "sensitivity function" that quantifies how small changes in the system parameters affect a particular design metric. The purpose of the present letter is to show that by exploiting the symplectic property [1] of Hamilton's equations, a sensitivity function can be defined that will aid in the design of charged particle beam sources used in accelerators and in beam-driven sources of coherent radiation such as free-electron lasers, klystrons, and traveling wave tubes. Further, we show that the sensitivity function can be computed by making a slight modification of the algorithm in the particle trajectory codes currently used to simulate beam sources.

   Normally, one would expect that to calculate a sensitivity function in an N-dimensional parameter space, N+1 computations would be required to determine the gradient of the metric in that space. Since N, the number of design parameters of interest, is generally very large, the required computational burden can be so large so as to rule out a full design optimization or sensitivity analysis. However, using adjoint techniques [2] this computational requirement can be reduced to as few as two computations – a 'base case' and a single, specially formulated perturbed case. The latter case is referred to as the 'adjoint' problem. The adjoint approach has previously been applied in circuit theory [3], electromagnetics [4], aerodynamics [5], plasma physics [6], as well as in other fields.

   The basic adjoint approach may be described as follows: Let the vector *X* represent the detailed state of the system with design parameters *B*; the dimension of *B* is N. For example, *X* could contain trajectory information for every particle in an electron gun and *B* could contain information on the locations and potentials of the electrode surfaces. We assume that small changes in *X* produced by small changes in *B* are related by a linear operator $A(X_0)$, which depends on the unperturbed system state $X_0$, that is, $A(X_0)\delta X = \delta B$. Now we usually don't need all of the information contained in $\delta X$, that is, complete, detailed knowledge of the change in system state is generally not required in the design process. Rather, what is important is the change in some particular performance metric(s), or figure(s) of merit *M* that depends on *X*. The change in *M* may be written, $\delta M = C^\dagger \delta X$ where $C^\dagger \equiv \partial M/\partial X$, and where the dagger symbol (†) indicates the adjoint (conjugate transpose) operator. The dimension of *M* is generally much smaller than *N*; in many cases of interest, the dimension of *M* is unity.

The key observation that provides the adjoint method its advantages is to note that the change in *M* can be evaluated without solving for $\delta X$. Specifically if we solve the adjoint equation $A^\dagger(X_0)\delta Y = C$ for $\delta Y$, then it follows that $\delta M = C^\dagger \cdot \delta X = \delta Y^\dagger \cdot \delta B$. Thus, a single inversion of the adjoint equation to find the sensitivity function $\delta Y^\dagger$ provides all the information needed to evaluate the change in the metric *M* for arbitrary changes in the parameters $\delta B$. We never need to compute $\delta X$.

In this Letter we develop a formalism using the adjoint method to characterize the sensitivity of electron beam parameters at the exit of an electron gun to variations in the gun design parameters such as electrode shape, position, and voltage, and magnetic field distribution and alignment. A 3D rendering of such an electron gun is shown in Fig. 1. While the example considered here is that of an electron gun, the method can be extended to other beam optics structures.

## 2. Adjoint Method for Electron Gun Design

Electron guns are generally designed using particle-in-cell simulation codes such as EGUN [7], DEMEOS [8], TRAK [9], UGUN [10], ARGUS [11], AVGUN [12], BOA [13], COCA [14] and MICHELLE [15]. These codes solve the equations of motion for electron trajectories in self-consistent electric and magnetic fields, starting at a cathode and ending at the exit of the gun where the beam is injected into some other structure. We denote the trajectory of particle *j* as $(\mathbf{x}_j(t), \mathbf{p}_j(t))$, where *x* and *p* are the position and momentum as functions of time since leaving the cathode; trajectory *j* is assumed to carry a given amount of steady current, $I_j$. The spatial distributions of charge and current density are

$$\left[\rho(\mathbf{x}), \mathbf{j}(\mathbf{x})\right] = \sum_j I_j \int_0^{T_j} dt \left[1, \mathbf{v}_j(t)\right] \delta(\mathbf{x} - \mathbf{x}_j(t)) \tag{1}$$

where $\mathbf{v}_j(t)$ is the velocity of particle *j*, and $T_j$ is the transit time of particle *j* through the system, that is, the time from emission to exit. These distributions are used as sources in Maxwell's equations to update the static fields,

$$\left[\nabla^2 \phi(\mathbf{x}), -\nabla \times \nabla \times \mathbf{A}(\mathbf{x})\right] = -\left[\rho(\mathbf{x})/\varepsilon_0, \mu_0 \mathbf{j}(\mathbf{x})\right] \tag{2}$$

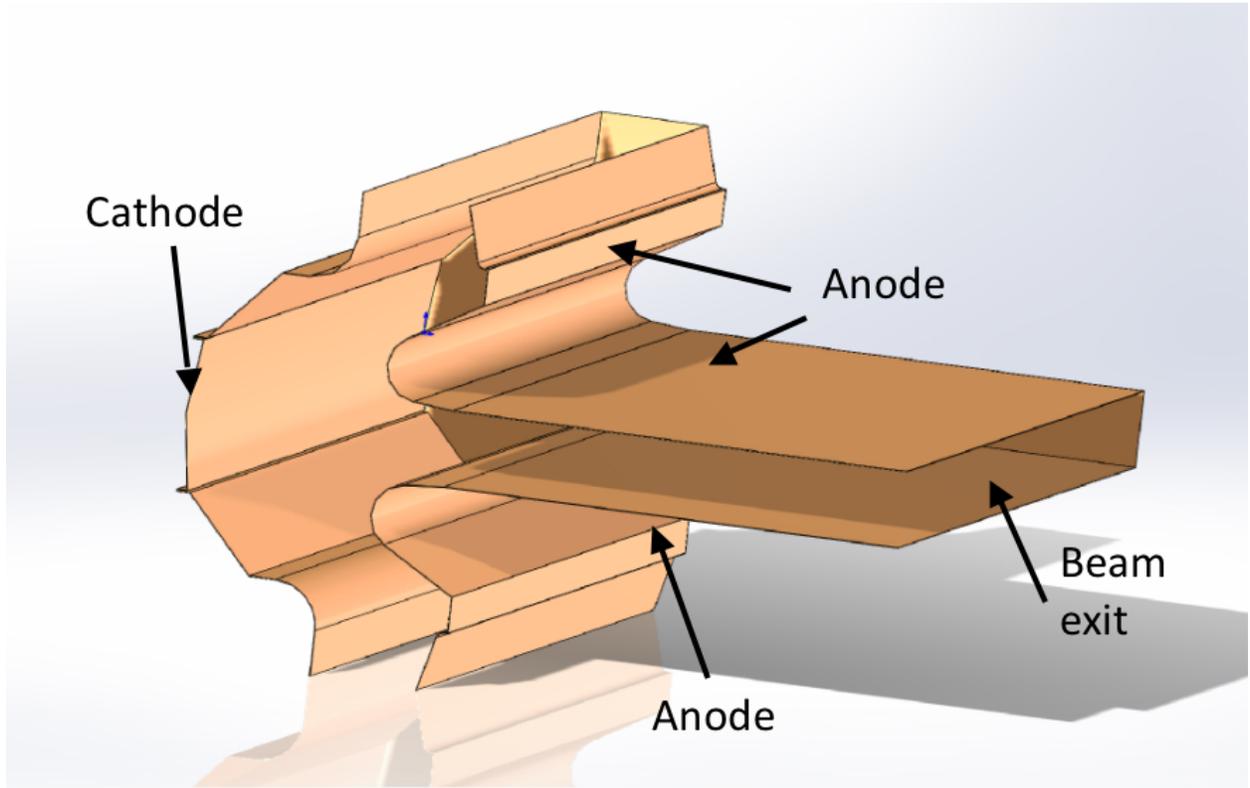

Fig.1: A 3D rendering of the electrodes of a sheet beam gun. The beam is drawn from the cathode on the left, accelerated and focused by the curved anode, and escapes through the beam exit on the right.

where $\phi(\mathbf{x})$ and $\mathbf{A}(\mathbf{x})$ are the electric and magnetic potentials, respectively. The process of launching particles, integrating trajectories, accumulating sources, and updating fields is repeated until a converged result is obtained. An example of a diode with planar symmetry is shown in Fig.1 and Fig.2. In this case the beam (shown in green in Fig. 2) is emitted from a cathode on the left, electrostatically focused by a pair of electrodes and extracted on the right through a drift space. The 2D calculation involved following the trajectories of 600 simulation particles through a domain consisting of a grid of 120,000 triangles. Typically, the process of computing trajectories, accumulating charge densities, and calculating self-consistent fields was iterated 200 times to reach a highly accurate converged solution.

Let us suppose that we have a reference design in which particle trajectories have been traced in combined electric and magnetic fields, and the fields have been calculated self consistently with the particle trajectories, as in Fig.2. This represents our base system state $X_0$. We consider a perturbation of this state such that trajectories $(\delta \mathbf{x}_j, \delta \mathbf{p}_j)$, charge and current densities $(\delta \rho, \delta \mathbf{j})$, and potentials $(\delta \phi, \delta \mathbf{A})$ all change. The equations of motion can be written in terms of a perturbed $q(\delta \Phi - \mathbf{v}_j \cdot \delta \mathbf{A})$ and an unperturbed Hamiltonian, $H(\mathbf{p}, \mathbf{x})$. The equations for the perturbed trajectories are

$$\frac{d\,\delta \mathbf{p}_j}{dt} = -\delta \mathbf{p}_j \cdot \frac{\partial^2 H}{\partial \mathbf{p}\partial \mathbf{x}} - \delta \mathbf{x}_j \cdot \frac{\partial^2 H}{\partial \mathbf{x}\partial \mathbf{x}} - q\frac{\partial}{\partial \mathbf{x}}\delta\Phi + q\frac{\partial}{\partial \mathbf{x}}\left(\mathbf{v}_j \cdot \delta\mathbf{A}\right) \qquad (3a)$$

$$\frac{d\,\delta \mathbf{x}_j}{dt} = \delta \mathbf{p}_j \cdot \frac{\partial^2 H}{\partial \mathbf{p}\partial \mathbf{p}} + \delta \mathbf{x}_j \cdot \frac{\partial^2 H}{\partial \mathbf{x}\partial \mathbf{p}} - q\frac{\partial}{\partial \mathbf{p}}\left(\mathbf{v}_j \cdot \delta\mathbf{A}\right). \qquad (3b)$$

Perturbed potentials satisfy linearized versions of (1) and (2) with boundary conditions to be specified.

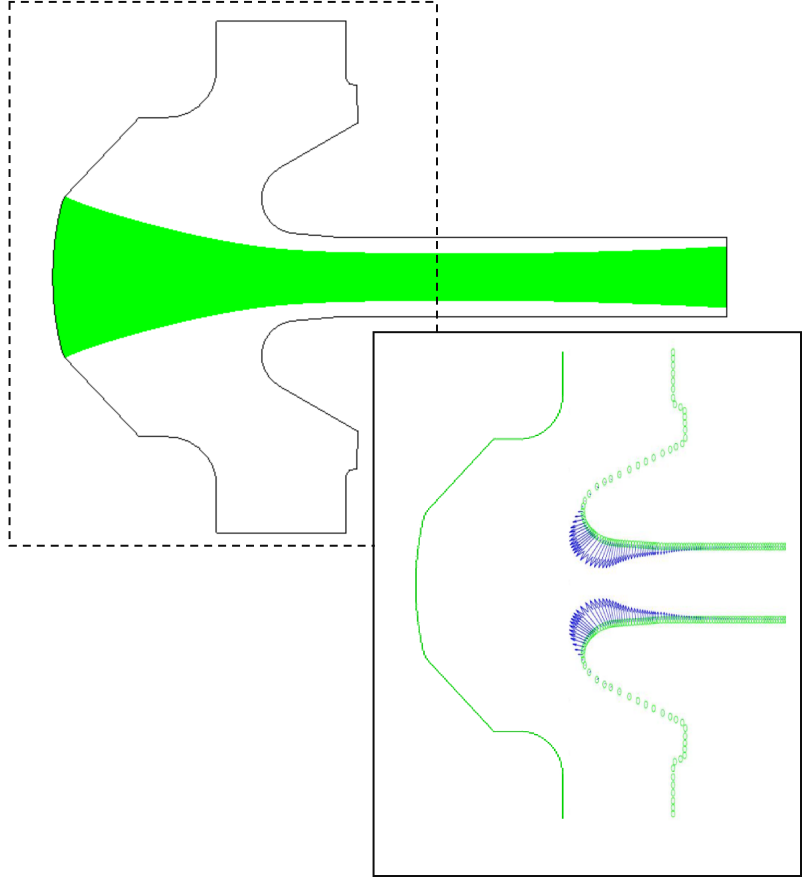

Fig. 2 Top 2D view of electron gun, beam is shown in green, cathode is on the left and anode on the right. Beam is extracted through a drift space. Inset shows the sensitivity function, in the form of arrows, for the RMS width of the beam at the exit.

We now consider two distinct perturbations: one ($\delta X$) caused by changing the parameters of the system ($\delta B$), and one that will be related to the adjoint solution ($\delta Y$). We refer to these as the "true" and "adjoint" perturbations and label them with superscripts $X$ and $Y$ respectively.

We dot Eqs. (3a) and (3b) for each of the perturbation types ($X$ or $Y$) with the change in position (3a) and momentum (3b) for the other perturbation type, subtract, multiply by the beamlet current for each trajectory, sum over trajectories, and integrate over time. The result is

$$\sum_j I_j \left( \delta \mathbf{p}_j^{(X)} \cdot \delta \mathbf{x}_j^{(Y)} - \delta \mathbf{p}_j^{(Y)} \cdot \delta \mathbf{x}_j^{(X)} \right)_0^{T_j} =$$

$$\sum_j I_j \int_0^{T_j} q dt \left\{ \left( \delta \mathbf{x}_j^{(X)} \cdot \frac{\partial}{\partial \mathbf{x}} + \delta \mathbf{p}_j^{(X)} \cdot \frac{\partial}{\partial \mathbf{p}} \right) \left( \delta \Phi^{(Y)} - \mathbf{v}_j \cdot \delta \mathbf{A}^{(Y)} \right) - (X \leftrightarrow Y) \right\}. \quad (4)$$

The left hand side of (4) is the current-weighted sum of the change, from cathode to a point near the exit, of the symplectic area defined by the pairs of trajectories. We will assume that the true trajectories ($X$) are unperturbed at the cathode, and consequently the left hand side of (4) only depends on perturbed trajectories near the exit. We will choose the perturbations of the adjoint trajectories ($Y$) near the exit such that the left side becomes the change in the metric $\delta M$ whose sensitivity function we seek.

Before doing this we note that the evaluation point is the time of flight of the unperturbed orbit, $T_j$. We want to replace this with an evaluation at, not just near, the exit of the device. The perturbed orbits will have different times of flight, given by $T_j + \delta T_j^{(X,Y)}$, to the exit of the domain. We can express perturbed quantities at the unperturbed time of flight in terms of their values at the exit of the domain and the rate of change of the corresponding unperturbed coordinate,

$$\left. (\delta \mathbf{q}_j^{(X,Y)}, \delta \mathbf{p}_j^{(X,Y)}) \right|_{T_j} = \left. (\delta \mathbf{q}_j^{(X,Y)}, \delta \mathbf{p}_j^{(X,Y)}) \right|_L - \delta T_j^{(X,Y)} \left. (\delta \dot{\mathbf{q}}_j, \delta \dot{\mathbf{p}}_j) \right|_{T_j}.$$

Using Hamilton's equations for the unperturbed trajectory the left hand side of (4) becomes

$$LHS = \sum_j I_j \left( \delta \mathbf{p}_j^{(X)} \cdot \delta \mathbf{x}_j^{(Y)} - \delta \mathbf{p}_j^{(Y)} \cdot \delta \mathbf{x}_j^{(X)} + \delta T_j^{(X)} \delta H_j^{(Y)} - \delta T_j^{(Y)} \delta H_j^{(X)} \right)_L,$$

where the $z$ components of the perturbed displacements now vanish. We note, the same result can be obtained by initially making a canonical transformation in which $z$ replaces $t$ as the independent variable. Here and below we will consider only perturbations that leave the Hamiltonians unchanged, i.e., $\delta H_j^{(X,Y)} = 0$.

The right side of (4) can be expressed in terms of the perturbed charge and current density. For example,

$$\sum_j I_j \int_0^{T_j} q dt \, \delta \mathbf{x}_j^{(X)} \cdot \frac{\partial}{\partial \mathbf{x}} \delta \Phi^{(Y)} = \int d^3 x \sum_j I_j \int_0^{T_j} dt \left( \delta(\mathbf{x} - \mathbf{x}_j - \delta \mathbf{x}_j^{(X)}) - \delta(\mathbf{x} - \mathbf{x}_j) \right) \delta \Phi^{(Y)}$$

$$= \int d^3 x \, \delta \rho^{(X)} q \delta \Phi^{(Y)}.$$

A similar relation applies to the vector potential terms. The result is

$$\sum_j I_j \left( \delta \mathbf{p}_j^{(X)} \cdot \delta \mathbf{x}_j^{(Y)} - \delta \mathbf{p}_j^{(Y)} \cdot \delta \mathbf{x}_j^{(X)} \right)_L = q \int d^3 x \left[ \delta \rho^{(X)} \delta \phi^{(Y)} - \delta \mathbf{j}^{(X)} \cdot \delta \mathbf{A}^{(Y)} - (X \leftrightarrow Y) \right] \quad (5)$$

The right hand side of (5) will become the sensitivity function. Its expression depends on the boundary conditions that apply to the solutions of the field equations (2).

We assume that electrostatic conditions are applied at the simulation boundary where either the potential or its normal derivative is specified for the true solution, and the perturbed potential $\delta\Phi^{(Y)}$ vanishes for the adjoint solution. For the magnetostatic fields, we assume the true system is perturbed by a change in current density in coils or in the magnetization of permanent magnets, $\delta \mathbf{j}_m(\mathbf{x})$; we also assume that all magnetic fields approach zero at infinity. We then apply Green's theorem to the integral on the right side of (5) to obtain

$$\sum_j \frac{I_j}{I}\left(\delta\mathbf{p}_j^{(X)}\cdot\delta\mathbf{x}_j^{(Y)} - \delta\mathbf{p}_j^{(Y)}\cdot\delta\mathbf{x}_j^{(X)}\right)_L = -\frac{q\varepsilon_0}{I}\int_B d^2x\,\delta\phi^{(X)}\mathbf{n}\cdot\nabla\delta\phi^{(Y)} + q\int d^3x\,\delta\mathbf{j}_m\cdot\delta\mathbf{A}^{(Y)} \quad (6)$$

Equation (6) is our central mathematical result. It is a form of Green's theorem extended to calculations using a mixture of fields and discrete particles. It states that the weighted sum of symplectic areas at the exit of the gun is given in terms of the perturbed electrostatic potential on the boundary and the perturbed current density creating the applied magnetic field; note that a displacement $\Delta(\mathbf{x})$ of the boundary is equivalent to a perturbed potential $\delta\Phi^{(X)}|_B = -\Delta(\mathbf{x})\cdot\nabla\Phi(\mathbf{x})$. These quantities are weighted by the sensitivity functions $\mathbf{n}\cdot\nabla\delta\phi^{(Y)}$ and $\delta\mathbf{A}^{(Y)}$; once these are known, Eq. (6) may be used to compute the symplectic areas for any perturbation $\delta\phi^{(X)}, \delta j_m^{(X)}$. We show below how the symplectic area may be related to a figure of merit of interest by suitable choice of the adjoint particle conditions $\left(\delta\mathbf{x}_j^{(Y)}, \delta\mathbf{p}_j^{(Y)}\right)$ at the exit plane. We pick the final coordinates, and integrate the trajectories backward in time through the gun, determining the self-consistent fields iteratively, exactly as in the forward integration for the unperturbed case.

We consider three example figures of merit for the gun of Fig. 1: a) the average vertical displacement of the beam, b) the RMS width of the beam, and c) the emittance of the beam. In the first case we pick as the final perturbed coordinates, $(\delta\mathbf{x}_j^{(Y)}=0, \delta\mathbf{p}_{\perp j}^{(Y)}=\lambda\mathbf{e}_\perp)_L$. That is, the perturbed momentum perturbation at the exit is a constant in the direction of the unit vector, $\hat{\mathbf{e}}_\perp$. In this case the left hand side of (6) becomes $-\lambda\sum_j I_j \hat{\mathbf{e}}_\perp\cdot\delta\mathbf{x}_j^{(X)}\Big|_L / I = -\lambda\hat{\mathbf{e}}_\perp\cdot\langle\delta\mathbf{x}_j^{(X)}\rangle$, where the angular brackets indicate a current weighted average. Thus, the left side of (6) is the average displacement in the direction of the unit vector. Both sides of Eq. (6) are proportional, in principle, to the constant $\lambda$, and it can then be divided out or set to unity to obtain the desired sensitivity. That would be the case if in practice we solved the linearized equations (3). However, this would require extensive programming to implement in the existing codes. Instead, we solve the nonlinear equations and select $\lambda$ to be sufficiently small, essentially taking a numerical derivative. This will be made clearer in the second example.

To determine the sensitivity of the RMS beam width we pick as final perturbed coordinates $(\delta\mathbf{x}_j^{(Y)}=0, \delta\mathbf{p}_{\perp j}^{(Y)}=\lambda\mathbf{x}_{\perp j})_L$ where $\lambda$ is again a constant, to be chosen so that the adjoint perturbations are sufficiently small. Then the left side of (6) becomes essentially the change in the RMS radius of the beam for the true solution,

$$\sum_j \frac{I_j}{I} \left( \delta \mathbf{p}_j^{(X)} \cdot \delta \mathbf{x}_j^{(Y)} - \delta \mathbf{p}_j^{(Y)} \cdot \delta \mathbf{x}_j^{(X)} \right)_{T_j} = -\lambda \sum_j \frac{I_j}{I} \delta \mathbf{x}_j^{(X)} \cdot \mathbf{x}_{\perp j} = -\frac{\lambda}{2} \delta \left\langle \left| \mathbf{x}_\perp \right|^2 \right\rangle \quad (7)$$

We thus can use the self-consistent adjoint fields as the sensitivity function for RMS beam radius.

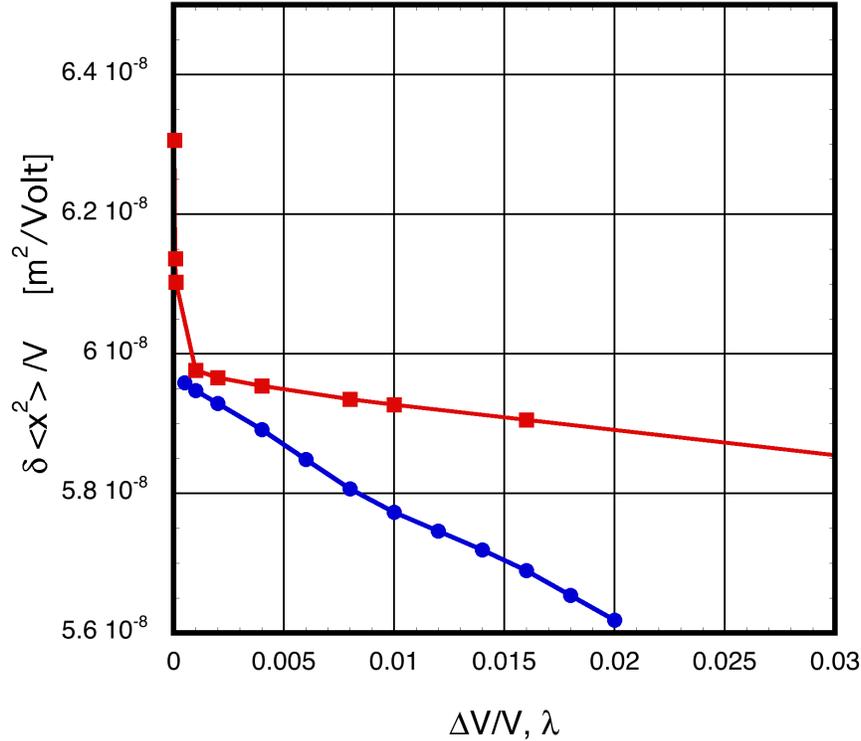

Fig. 3. Change in RMS beam width squared per volt change in anode potential calculated two ways. Shown in blue circles is the change in width calculated directly plotted versus the relative change in anode voltage. Shown in red squares is the change in width obtained from the adjoint calculation plotted versus the numerical factor controlling the linearization of the equations of motion for the adjoint calculation..

To test this approach we will compute the change in RMS beam thickness in two ways, for the device in Fig. 1, due to a change in potential on the anode. The first way is to directly simulate the device with a range in applied voltages. This will allow us to determine the range in voltage in which the changes in RMS thickness are linear in the change in applied voltage. Shown in Fig. 3 with round blue symbols is the change in mean squared beam width divided by the change in voltage as a function of the relative change in voltage. The line appears to be straight as would be expected for a numerically obtained derivative and shows an intercept of about 5.96 x $10^{-8}$ m$^2$/V as change in voltage decreases to zero. This is to be compared with the value obtained from the adjoint formulation shown by the curve with the red squares in Fig. 3. Here we have computed the sensitivity function in the form of $-n \cdot \nabla \phi^{(Y)} / \lambda$ obtained from the backwards integration of the particles with final conditions specified according to (7). The sensitivity function is defined on the boundaries of the simulation domain and is shown as arrows on Fig. 2. This emphasizes that the sensitivity function can be used to determine the change in mean squared beam thickness for any change in anode shape or potential, not just the uniform change in potential considered here. We compute the expected change in the mean square beam width for a one volt change in anode potential by doing the surface integral in (6) numerically. The results are then plotted in Fig. 3 as a function of the constant $\lambda$. For $\lambda > 0.001$ the curve shows expected linear behavior with an intercept that is within 1% of the value obtained from the true calculations. For extremely small values of $\lambda < 0.001$ the accuracy of the numerical differentiation is lost due to a number of limitations associated with numerical differentiation.

As a final example, we may determine the sensitivity of the beam emittance to changes in electrode placement. To do so, we consider changes in the quantity

$$\varepsilon^2 = \langle |\mathbf{p}_\perp|^2 \rangle \langle |\mathbf{x}_\perp|^2 \rangle - \langle \mathbf{p}_\perp \cdot \mathbf{x}_\perp \rangle^2, \tag{8}$$

where we have assumed that the average of the transverse positions and momenta vanish. The change of the emittance is then given by

$$\delta\varepsilon^2/2 = \langle \delta\mathbf{p}_\perp^X \cdot \mathbf{p}_\perp \rangle \langle |\mathbf{x}_\perp|^2 \rangle + \langle |\mathbf{p}_\perp|^2 \rangle \langle \delta\mathbf{x}_\perp^X \cdot \mathbf{x}_\perp \rangle - \langle \mathbf{p}_\perp \cdot \mathbf{x}_\perp \rangle \langle \delta\mathbf{p}_\perp^X \cdot \mathbf{x}_\perp + \delta\mathbf{x}_\perp^X \cdot \mathbf{p}_\perp \rangle. \tag{9}$$

Thus if we choose

$$\delta\mathbf{x}_\perp^Y = \lambda \left( \mathbf{p}_\perp \langle |\mathbf{x}_\perp|^2 \rangle - \mathbf{x}_\perp \langle \mathbf{p}_\perp \cdot \mathbf{x}_\perp \rangle \right) \tag{10a}$$

and

$$\delta\mathbf{p}_\perp^Y = \lambda \left( -\mathbf{x}_\perp \langle |\mathbf{p}_\perp|^2 \rangle + \mathbf{p}_\perp \langle \mathbf{p}_\perp \cdot \mathbf{x}_\perp \rangle \right), \tag{10b}$$

then the left hand side of (6) becomes $\lambda\delta\varepsilon^2/2$ and so the change in beam emittance due to changes in electrode potential or position may be computed using Eq.(6).

### 3. Summary

A sensitivity function for design of electron beam optics has been introduced. The sensitivity function, which gives the change in some metric for arbitrary changes in electrode potentials and shapes, and arbitrary changes in coil or magnet locations, can be calculated with a few runs of the codes currently used to design electron guns. While the present situation is static, the approach can be generalized to time varying situations as well.

### 4. Acknowledgment

This work was supported by DARPA contract HR0011-16-C-0080 with Leidos, Inc. The views, opinions and/or findings expressed are those of the authors and should not be interpreted as representing the official views or policies of the Department of Defense or the U.S. Government. Approved for Public Release, Distribution Unlimited.

---

[i] University of Maryland, College Park, MD 20742